\documentclass[12pt,preprint]{aastex}

\def\rf#1 {{\sl ({#1})}}
\begin{document}
\title{Interstellar Scintillation Velocities of a Relativistic Binary PSR 
B1534+12 and Three Other Millisecond Pulsars}

\author{Slavko Bogdanov}
\affil{Department of Astronomy and Astrophysics, Penn State University \\
University Park, PA 16802, USA \\
bogdanov@astro.psu.edu}
\author{Ma\l gorzata Pruszy\' nska and Wojciech Lewandowski}
\affil{Toru\'n Centre for Astronomy, Nicolaus Copernicus University \\
ul. Gagarina 11, 87-100 Toru\' n, Poland\\
mare@astro.uni.torun.pl, boe@astro.uni.torun.pl}
\author{Alex Wolszczan}
\affil{Department of Astronomy and Astrophysics, Penn State University \\
University Park, PA 16802, USA \\
Toru\'n Centre for Astronomy, Nicolaus Copernicus University \\
ul. Gagarina 11, 87-100 Toru\' n, Poland\\
alex@astro.psu.edu}

\begin{abstract}

We present interstellar scintillation velocity measurements for four 
millisecond pulsars obtained from long-term monitoring observations 
with the Arecibo radio telescope at 430 MHz.
We also derive explicit expressions that relate the measured scintillation 
velocity to the effective transverse velocity responsible for the motion of 
the diffraction pattern for both binary and solitary pulsars.
For B1257+12, B1534+12, J1640+2224, and J1713+0747 we derive
ISS velocity estimates of $197\pm57$, $192\pm63$, $38\pm8$, 
and $82\pm16$ km s$^{-1}$, respectively. 
These values are in good agreement with proper motion measurements 
for the four pulsars. For a relativistic binary pulsar PSR B1534+12, we 
use the ISS velocity dependence on orbital phase to determine the longitude 
of the ascending node $\Omega$ of the pulsar's orbit and to 
derive an estimate of the effective scattering screen location.
The two possible values of $\Omega$ are $70\pm20$ and 
$290\pm20$ degrees and the approximate screen location is
$630 \pm 200$ pc with the assumed pulsar distance of $1.1$ kpc.

\end{abstract}

\keywords{pulsars --- interstellar medium --- pulsars: relativistic binaries ---  pulsars: individual (PSR B1257+12, PSR B1534+12, PSR J1640+2224, PSR J1713+0747)}

\section{INTRODUCTION}

Interstellar scintillation (ISS) of pulsars
is caused by diffraction and refraction of radio waves from
electron density irregularities in the interstellar plasma \citep{Ric90}. 
Measurements of the ISS related modifications
of the pulsar signal, such as the angular source broadening,
temporal broadening of pulses and their intensity variations in
time and frequency, have been used to study the small-scale structure of 
the interstellar plasma \citep*[e.g.][]{Cor85,Lamb00}.

Pulsars are fast moving objects with mean velocities
of $\sim$460 km s$^{-1}$ \citep*{Lyne94} and
$\sim$85 km s$^{-1}$ \citep*{Tosc99} for the normal and the millisecond 
pulsar populations, respectively. Therefore, measurements of a net speed of 
the interstellar diffraction pattern derived from
decorrelation scales of pulse intensity 
variations in time and frequency are dominated by the pulsar motion and 
can be used to estimate a transverse component of the pulsar velocity 
in space. In fact, the ISS technique has been routinely
used as one of the means to measure pulsar velocities  
\citep*[e.g.][]{Lyne82,Cor86,Gup94}, in addition to the more direct methods 
of pulse timing \citep[e.g.][]{Tosc99}
and radio interferometry \citep[e.g.][]{Fom97}. As the latter
two techniques are limited to low timing noise, bright, nearby
objects, the ISS observations provide a very helpful,
additional way to obtain velocity estimates for a relatively
large number of sources. Such measurements, in combination with
simulation studies \citep{Tau96,Cor97}
are important for global characterization of the galactic population of 
neutron stars.

The ISS velocity measurements, when combined with timing or 
interferometric astrometry data, provide useful constraints on the 
distribution of scattering material along the pulsar line of sight 
and provide a way to verify a reliability of the ISS method itself 
\citep{Cor98,Des98}. Recent analyses 
of a growing body of such measurements 
have shown that, despite their obvious model dependence, the ISS velocities 
are remarkably accurate \citep*{Gup95,John98,Nic01}.

In principle, the ISS observations of binary pulsars can be used to track 
their orbital velocities projected onto the plane of the sky and to deduce 
some orbital parameters that would be impossible to measure in any other 
way \citep*{Lyne84,Dew88}. This is particularly true for relativistic
binaries, in which case the orbital velocity variations are exceptionally
large and may dominate the ISS measurements.

In this paper, we present ISS velocity measurements of four 
millisecond pulsars obtained from long-term monitoring observations with 
the 305-m Arecibo radio telescope. PSR J1640+2224 \citep*{Fos95} and 
PSR J1713+0747 \citep*{Fos93} are binary millisecond pulsars
in long period, nearly circular orbits with white dwarf companions. 
PSR B1257+12 is a fast moving solitary millisecond pulsar 
with a system of at least three planets around it \citep*{Wol92,Wol94}. 
Finally, PSR B1534+12 is a relativistic 10-hour binary pulsar in orbit 
with another neutron star \citep*{Wol91,Stairs98}. Here, we focus our 
attention on PSR B1534+12 and the measurements of changes in its ISS 
velocity caused by orbital motion. We also verify the  
earlier ISS speed measurements for PSR B1257+12 and PSR J1713+0747 
\citep{John98,Goth00} and provide the first such measurement for 
PSR J1640+2224.

\section{OBSERVATIONS AND ANALYSIS OF THE DYNAMIC SPECTRA}

The four pulsars discussed in this paper have been regularly monitored 
with the Arecibo radio telescope to measure their pulse arrival times. The 
most recent results of timing modeling of these objects will be presented 
elsewhere. We have reprocessed the timing measurements made at 
multiple epochs between 1994 and 2001 at 430 MHz to obtain the dynamic 
spectra of pulsar scintillations and analyzed them to estimate the ISS 
velocities of the pulsars themselves. 

Typically, PSR B1257+12, PSR J1640+2224 and PSR J1713+0747 were 
observed for 45-120 min. in a series of 3 min. integrations carried out 
synchronously with the Doppler-corrected pulsar period. 
In the case of PSR B1534+12, the dynamic spectra were analyzed in 
45-60 min. sections to avoid excessive averaging over changes in the 
frequency and time decorrelation scales induced by the pulsar's orbital 
motion. Similarly, to eliminate biases in the decorrelation scales 
introduced by refractive 
scintillation \citep*{Bhat99} the dynamic spectra strongly affected by 
drifts of spectral features and quasiperiodic fringing were excluded from 
the analysis.

The left- and right-hand circularly polarized pulsar
signals were fed into the Penn State Pulsar Machine 
(PSPM), which is a 2 $\times$ 128 $\times$ 60 kHz channel, computer 
controlled, on-line pulsar processor. After detection, the dual-channel 
signals were added together, smoothed, 4-bit quantized and pulse phase 
averaged by the analysis computer in each of the 128 frequency channels. 
In the off-line processing, the 128 $\times$ 60 kHz point
spectra of the pulsar signal were formed from the ON-pulse and OFF-pulse 
level measurements in each frequency channel by calculating the 
corresponding (ON-OFF)/OFF spectral intensities. A dynamic spectrum of 
the pulsar intensity variations in time and frequency was then 
constructed from a set of 3-minute averaged, contiguous spectra for 
each object and each observing session. 

Examples of the dynamic spectra of the four pulsars are shown in 
Figures 1 and 2. Also shown are the two-dimensional autocorrelation 
functions of these spectra,
which were used to measure the decorrelation scales of 
pulsar intensity variations, $\Delta t_d$ and 
$\Delta\nu_d$. Following the established convention, $\Delta t_d$ and 
$\Delta\nu_d$ were measured as the respective $1/e$ widths and half-widths 
of elliptical Gaussians least-squares fitted to the autocorrelation 
functions. The decorrelation widths for PSR B1257+12, PSR B1534+12, 
PSR J1640+2224 and PSR J1713+0747, averaged over all observing epochs are 
listed in Table 1.

The errors of $\Delta t_d$ and $\Delta\nu_d$ measurements consist of 
statistical uncertainty due to a finite number of ``scintles'' in the 
dynamic spectra, the signal strength, and the long-term changes of these 
parameters caused by refractive effects \citep{Bhat99}. We found 
that the refraction induced variations in $\Delta t_d$ and $\Delta\nu_d$ 
occurring on the time scales from months to years were a dominant source of 
error for all the four pulsars discussed here. Consequently, the errors 
quoted in Table 1 were calculated from the scatter in the estimates of the 
decorrelation widths over the entire time span of observations.

\section{THE ISS VELOCITY ANALYSIS}

The ISS velocities of pulsars can be calculated from their dynamic 
spectra, because the two observables, the characteristic decorrelation
scales of pulse intensity in time, $\Delta t_d$, and in frequency, 
$\Delta\nu_d$ are directly related to the net speed of a diffractive 
scintillation pattern \citep{Lyne82}. For $\Delta t_d$ and $\Delta\nu_d$ 
measured in seconds and megahertz, respectively, the velocity $V_{ISS}$ 
(km s$^{-1}$) is derived from the equation \citep{Gup94}:
\begin{equation}
V_{ISS}=3.85 \times 10^4 \frac{\sqrt{\Delta\nu_d d x}}{\nu\Delta t_d},
\end{equation}
where $x$ is the ratio of the distances of the scattering screen to the 
observer and to the pulsar, $d$ is the pulsar distance (in kpc), and $\nu$ 
is the observing frequency (in GHz). The numerical constant
is valid for the assumed homogeneous turbulent medium characterized by
a Kolmogorov turbulence spectrum.

In principle, the observed ISS velocity of a pulsar represents 
the net effect of the pulsar's motion, the Earth's orbital motion around 
the Sun, and the motion of the scattering medium. Thus, $V_{ISS}$
can be written as the magnitude of the vector 
sum of all the velocities contributing to it \citep{Gup94}:
\begin{equation}
V_{ISS}=|x\mbox{\boldmath$ V_o$}+x\mbox{\boldmath$ V_{pm}$}
+\mbox{\boldmath$ V_{\oplus}$}-(1+x)\mbox{\boldmath$ V_{ism}$}|_{\perp}
\end{equation}
where, for binary pulsars, \mbox{\boldmath $V_o$} is the pulsar's orbital 
velocity, \mbox{\boldmath $V_{pm}$} is its proper motion velocity, 
\mbox{\boldmath $V_{\oplus}$} is the 
velocity of the Earth, and \mbox{\boldmath $V_{ism}$} is the velocity of the 
scattering screen. The subscript $\perp$ indicates that the only 
contributions to the ISS velocity come from the components of the velocity 
vectors perpendicular to the line of sight.
Choosing the origin of a coordinate system to coincide with the center
of mass of a binary with its x-y plane tangent to the sky, with the x-axis
parallel to the orbit's line of nodes and the z-axis pointing from the solar 
system barycenter to that of the binary system, $V_{ISS}$ can be expressed 
in terms of the projections of the component velocity vectors onto the 
plane of the sky:
\begin{eqnarray}
V_{ISS}^2= & (xV_{o,x}+xV_{pm,x}+V_{\oplus,x})^2 \nonumber \\
 & +(xV_{o,y}+xV_{pm,y}+V_{\oplus,y})^2
\end{eqnarray}
In this expression we have assumed that the contribution of $V_{ism}$ 
to $V_{ISS}$ is negligible \citep{Gup95}. The component velocities in 
Eqn. (3) are defined as follows. The orbital velocity components are:
\begin{eqnarray}
V_{o,x}=\xi
 & \lbrace[e\sin\phi\cos\phi-(1+e\cos\phi)\sin\phi]\cos\omega \nonumber \\
 & -[e\sin\phi\sin\phi+(1+e\cos\phi)\cos\phi]\sin\omega\rbrace
\end{eqnarray}
\begin{eqnarray}
V_{o,y}=\xi
 & \lbrace[e\sin\phi\cos\phi-(1+e\cos\phi)\sin\phi]\sin\omega \nonumber \\
 & +[e\sin\phi\sin\phi+(1+e\cos\phi)\cos\phi]\cos\omega\rbrace(\pm\cos i)\end{eqnarray}
where $\xi = (2\pi a \sin i) / (P_b\sqrt{1-e^2}\sin i)$, $a$ is the 
semi-major axis, $e$ is the eccentricity, $P_b$ is the orbital period, 
$\phi$ is the orbital phase, $\omega$ is the longitude of periastron, 
and $i$ is the orbital inclination. 
The two possible signs of the $\cos i$ term arise from the fact that
we have no way of determining which of the two nodes of the orbit is the 
ascending node so both cases need to be considered. The proper motion 
components are defined by the familiar relationships:
\begin{equation}
V_{pm,x}=4.74d(\mu_{\alpha}\sin\Omega+\mu_{\delta}\cos\Omega)
\end{equation}
\begin{equation}
V_{pm,y}=4.74d(-\mu_{\alpha}\cos\Omega+\mu_{\delta}\sin\Omega)
\end{equation}
where $\Omega$ is the longitude of the ascending node of the orbit, defined 
as the position angle (reckoned from north through east) of the line of nodes 
measured in the plane tangent to the sky, 
$\mu_{\alpha}$ and $\mu_{\beta}$ are the proper motions in right ascension
and declination, respectively, in units of mas yr$^{-1}$, and the distance 
$d$ is given in kpc. Finally, the Earth velocity components are given by:
\begin{eqnarray}
V_{\oplus,x} = & 1.73\times10^{-6} & 
[\dot{X}(-\cos\Omega\sin\delta\cos\alpha-\sin\Omega\sin\alpha) \nonumber \\
& & +\dot{Y}(-\cos\Omega\sin\delta\sin\alpha+\sin\Omega\cos\alpha) \\
& & +\dot{Z}(\cos\Omega\cos\delta)] \nonumber
\end{eqnarray}
\begin{eqnarray}
V_{\oplus,y} = & 1.73\times10^{-6} & 
[\dot{X}(-\sin\Omega\sin\delta\cos\alpha+\cos\Omega\sin\alpha) \nonumber \\
& & +\dot{Y}(-\sin\Omega\sin\delta\sin\alpha-\cos\Omega\cos\alpha) \\
& & +\dot{Z}(\sin\Omega\cos\delta)] \nonumber
\end{eqnarray}
where $\alpha$ and $\delta$ are the pulsar coordinates and the parameters 
$\dot X$, $\dot Y$, and $\dot Z$ are the components of Earth's velocity in the
Solar System Barycenter coordinate system expressed in units of AU d$^{-1}$.

Expression (3) conveniently reduces to one for a solitary pulsar by 
simply neglecting the orbital motion terms and setting $\Omega$ equal 
to zero:
\begin{eqnarray}
V_{ISS}^2 = & 
\lbrace x(4.74d\mu_{\delta}) & 
+1.73\times10^{-6}[\dot{X}(-\sin\delta\cos\alpha)+\dot{Y}(-\sin\delta
\sin\alpha)+\dot{Z}\cos\delta] \rbrace^2 \nonumber \\
 & + \lbrace x(-4.74d\mu_{\alpha}) &  +1.73\times10^{-6}
[\dot{X}\sin\alpha-\dot{Y}\cos\alpha] \rbrace^2
\end{eqnarray} 
In what follows,
Eq. (1) is used to derive the observed ISS velocities from ISS 
measurements, whereas Eq. (3) or Eq. (10) give the expected ISS speeds 
from the known pulsar proper motion and Earth velocities.

\section{ISS VELOCITY AND ORBITAL MOTION OF PSR B1534+12}

\citet{John98}, \citet{Goth00}, and 
\citet{Lomm01} have measured the ISS velocity for this pulsar to be 
190 km s$^{-1}$, 191 km s$^{-1}$, and 149 km s$^{-1}$, respectively. In 
principle, these estimates can be significantly biased by a varying 
transverse component of the pulsar's 
orbital velocity, whose semi-amplitude exceeds $\sim$100 km s$^{-1}$. Here we
have performed a rigorous analysis of this effect by taking into account 
all the motions contributing significantly to the observed ISS velocity of 
PSR B1534+12, as well as the dependence of the ISS velocity on the pulsar's 
orbital phase.

Compared to the results of
similar analyses carried out for the binary pulsars PSR B0655+64 by 
\citet{Lyne84} and PSR B1855+09 and PSR B1913+16 by \citet{Dew88},
the ISS velocities for PSR B1534+12 shown in Fig. 3 
exhibit a much more pronounced dependence 
on orbital phase and allow an unambiguous modeling of this effect
with the aid of Eqs. (3-9). In fact, the problem can be conveniently 
parametrized in terms of only two unknowns,
the longitude of the ascending node $\Omega$ and the screen location $x$,
with the Keplerian orbital parameters and proper motion of the pulsar  
already determined from long-term timing measurements 
\citep[and references therein]{Stairs98}. For the pulsar distance we 
have adopted a value of $d = 1.1 \pm 0.2$ kpc derived by \citet{Stairs98} 
on the assumption that General Relativity (GR) is the correct theory 
of gravity. 

Although the longitude of periastron $\omega$ of the pulsar's orbit 
advances at a rate of 1.75 deg yr$^{-1}$, it does not change fast enough 
to have a significant effect on our analysis. Consequently, we have used 
the value of $\omega$ at the mid-point of the actual observing 
period. Finally, because the transverse component of the Earth's 
velocity varies and its direction relative to the pulsar's orbital motion 
cannot be determined without knowing $\Omega$, we have ignored the 
$V_{\oplus}$ terms in Eq. (3) and included them in the uncertainties of 
the $V_{ISS}$ measurements. 

Since both the model (Eq. 3) and Eq. (1) 
include a dependence on the parameter $x$, the best fit values of
$\Omega$ and $x$ were obtained by manually searching the $\chi^{2}$-space for 
the minimum. The fit yielded $\Omega_1=70 \pm 20$ and 
$\Omega_2=290 \pm 20$ degrees for the longitude of the ascending node and 
$x=1.3 \pm 0.4$, which corresponds to a screen located at a distance of 
$d_s=630 \pm 200$ pc from the Earth. Fig. 3 shows the ISS velocity
measurements of PSR B1534+12 and the best fitting curve given by Eq. (3)
plotted against orbital phase.

>From our ISS velocity measurements of PSR B1534+12 it is also possible 
to give an estimate of the pulsar's transverse systemic velocity. 
Since the transverse component of the pulsar's orbital velocity 
never vanishes the best we can do is to consider only the measurements made
around orbital phases 0.2 and 0.7 at which it reaches the minimum value.
Using these measurements we 
obtain a value of 192 km s$^{-1}$, which is an overestimate of the actual 
proper motion due to contributions from the pulsar's orbital motion as well
as the location of a scattering screen. Nonetheless, 
this value is remarkably close to those published by \citet{John98} and 
\citet{Goth00}.

\section{ISS VELOCITIES OF PSR B1257+12, PSR J1640+2224, AND PSR J1713+0747}

Scintillation velocities of PSR B1257+12 were calculated using the distance
based on the \citet{Tay93} model of galactic electron distribution 
(the TC model hereafter). For the pulsar distance of 0.62 kpc, we obtained an 
average value of $V_{ISS} =197 \pm 57$ km s$^{-1}$. The ISS velocity estimate 
of 225 km s$^{-1}$ using the same distance has been derived by 
\citet{Goth00}. We find these two estimates to be remarkably 
similar given the large, long-term variations of the 
ISS velocity observed for this pulsar. The measured velocity variations 
ranging from 117 km s$^{-1}$ to 290 km s$^{-1}$ on a time scale of $\sim$2 
years
are consistent with modifications of the scintillation decorrelation scales 
by the refractive scintillation. We have found the ISS   
velocity of this pulsar to be consistently lower 
than the effective transverse velocity of $\sim$278 km s$^{-1}$, computed 
using the proper motion values given by \citet{Wol00}, with the 
average ratio $V_{ISS}/V_{eff}=0.67$. 
This difference can be plausibly accounted for by assuming that the
scattering screen is located closer to the observer than to the 
pulsar ($x<1$). It is also likely that the distance used is an
underestimate of the true distance to the pulsar. 

Our measurement of $V_{ISS}$ of 38 $\pm$ 8 km s$^{-1}$ for PSR J1640+2224
represents the first ISS velocity estimate for this binary millisecond pulsar.
In the calculations we adopted the TC model distance of 1.18 kpc.
In comparing this velocity with the value of 75 km s$^{-1}$ obtained from 
Eq. (10), computed with proper motion measurements provided by 
\citet{Wol00}, we have ignored
the transverse component of the pulsar's orbital velocity, because it is 
unlikely to be greater than $10$ km s$^{-1}$. On the other hand, 
the transverse component of the Earth's 
orbital velocity accounts for as much as 40\% of the effective ISS velocity 
of the pulsar and it must be included in the analysis. 
Since the ISS velocity for PSR J1640+2224 is 
lower than the effective transverse velocity with the ratio 
$V_{ISS}/V_{eff}=0.52$, it is likely that the effective location of the 
scattering screen is closer to the observer than to the pulsar. However,
this discrepancy may also mean that the distance of 1.18 kpc we used may
be an underestimate of the actual pulsar distance.  

ISS velocity estimates of 29 km s$^{-1}$ and 118 $\pm$ 40 km s$^{-1}$ for 
PSR J1713+0747 have already been published by \citet{John98}, using 
a distance of 0.89 kpc, and by \citet{Lomm01}, respectively. Even if 
corrected for the TC distance of 1.1 kpc used here, the value given by 
Johnston et al. is still significantly lower than our velocity estimate 
of 82 $\pm$ 16 km s$^{-1}$ and that of \citet{Lomm01}. 
In absence of observations made at sufficiently 
similar epochs, the source of large discrepancies between the three 
values remains uncertain, although the refraction induced $V_{ISS}$ 
variability provides a possible explanation. Since the transverse component 
of the orbital velocity of PSR J1713+0747 does not contribute more than 
$\sim 15$ km s$^{-1}$ to the pulsar's ISS velocity, we included it in the 
effective uncertainty of the measurement and used Eq. (10) instead of 
Eq. (3).  With a net effective velocity of 52 km s$^{-1}$ 
for this pulsar, calculated using the proper motion values given by 
\citet*{Cam94}, our measurements of the ISS velocity are
consistently larger than the effective transverse velocity. In this case, 
the ratio $V_{ISS}/V_{eff}$ = 1.71 suggests either the presence of an 
asymmetrically located scattering screen which is closer to the pulsar than 
to the observer (i.e. $x>1$) or that the pulsar distance is overestimated.

\section{DISCUSSION}
We have presented measurements of interstellar scintillation 
velocities for four pulsars together with a complete set of relationships 
that should be used to compare these velocities with the proper motion 
measurements. Our $V_{ISS}$ estimates are generally consistent with the 
corresponding proper motion based values derived from timing measurements of 
these objects. As shown by the results of similar analyses conducted in 
the past \citep[e.g][]{John98,Goth00}, the discrepancies 
between these two kinds of the transverse pulsar velocity measurements can 
be accounted for by either an asymmetric location of the effective 
scatterer along the line of sight, or by an inaccurate estimate
of the pulsar distance. In principle, these two factors can be decoupled by 
the analysis that involves the angular and temporal broadening data, in 
addition to the scintillation and proper motion measurements of a 
pulsar \citep{Cor98,Des98}. It is likely, that these ``hybrid'' methods 
will offer an important means to study a line-of-sight distribution of the 
scattering material, as timing and astrometric proper 
motion measurements for many more pulsars become available.

Our analysis of the relativistic binary pulsar PSR B1534+12 provides
the best demonstration so far that the changes of a transverse component 
of the pulsar orbital velocity can be unambigously detected in the ISS 
velocity data. The most important practical consequence of this detection 
is the measurement of the longitude of the ascending node $\Omega$ of the 
binary orbit. This parameter cannot be determined from pulsar timing, 
except in very special circumstances \citep{Kop96,van01}.
One of the instances where the knowledge of $\Omega$ is important is the 
effect of aberration caused by the relativistic orbital motion of a 
pulsar \citep{Dam92}. In this case, $\Omega$ has to be
taken into account in the determination of aberration-induced variations
of the orbital phase dependent offset of the linear polarization angle of 
the pulse profile \citep[Eqs. 2.16-2.24]{Dam92}. The value of  
$\Omega$ is also needed for a precise measurement of a proper motion 
related rate of change of the projected size of the pulsar orbit  
\citep[and references therein]{Kop96}: 
\begin{equation}
\delta \dot{x}= 1.54 \times 10^{-16} x \cot i (-\mu_{\alpha} \sin \Omega + \mu_{\delta} \cos \Omega)
\end{equation}
\noindent
where, $x$ is the size of the projected semi-major axis 
expressed in seconds. From this equation we obtain values of 
$\delta \dot{x}_1= -9.43\times 10^{-16}$ s s$^{-1}$ and
$\delta \dot{x}_2 = -1.25 \times 10^{-15}$ s s$^{-1}$, corresponding to
$\Omega_1$ and $\Omega_2$, respectively. With the currently available 
timing data for PSR B1534+12, this effect is not yet measurable.

Aside from the ambiguity in $\Omega$ resulting from the unknown direction 
of the pulsar's orbital motion, the two values of this parameter obtained 
from our analysis are very robust in the sense that small changes of 
$\Omega$ lead to the correspondingly large changes of the shape of the 
$V_{ISS}$ curve (Fig. 3). On the other hand, the screen location
is less well determined, because of the uncertain estimate of the pulsar 
distance. However, if the GR derived distance to PSR B1534+12 is reliable, 
it is probably safe to assume that the screen is indeed located closer to 
the pulsar than to the observer.

We would like to thank Ingrid Stairs for her continuing help with the 
PSPM observations of PSR B1534+12 and Maciej Konacki for insightful 
discussions. This work was supported by a National Science Foundation
grant AST-9988217 (SB, AW) and a Polish Committee for Scientific
Research grant 2-P03D-006-16 (WL, MP, AW).
The Arecibo Observatory is part of the National Astronomy and 
Ionosphere Center, which is operated by Cornell University under a 
cooperative agreement with the NSF.

\clearpage

%
% FIGURE CAPTIONS
%

\figcaption[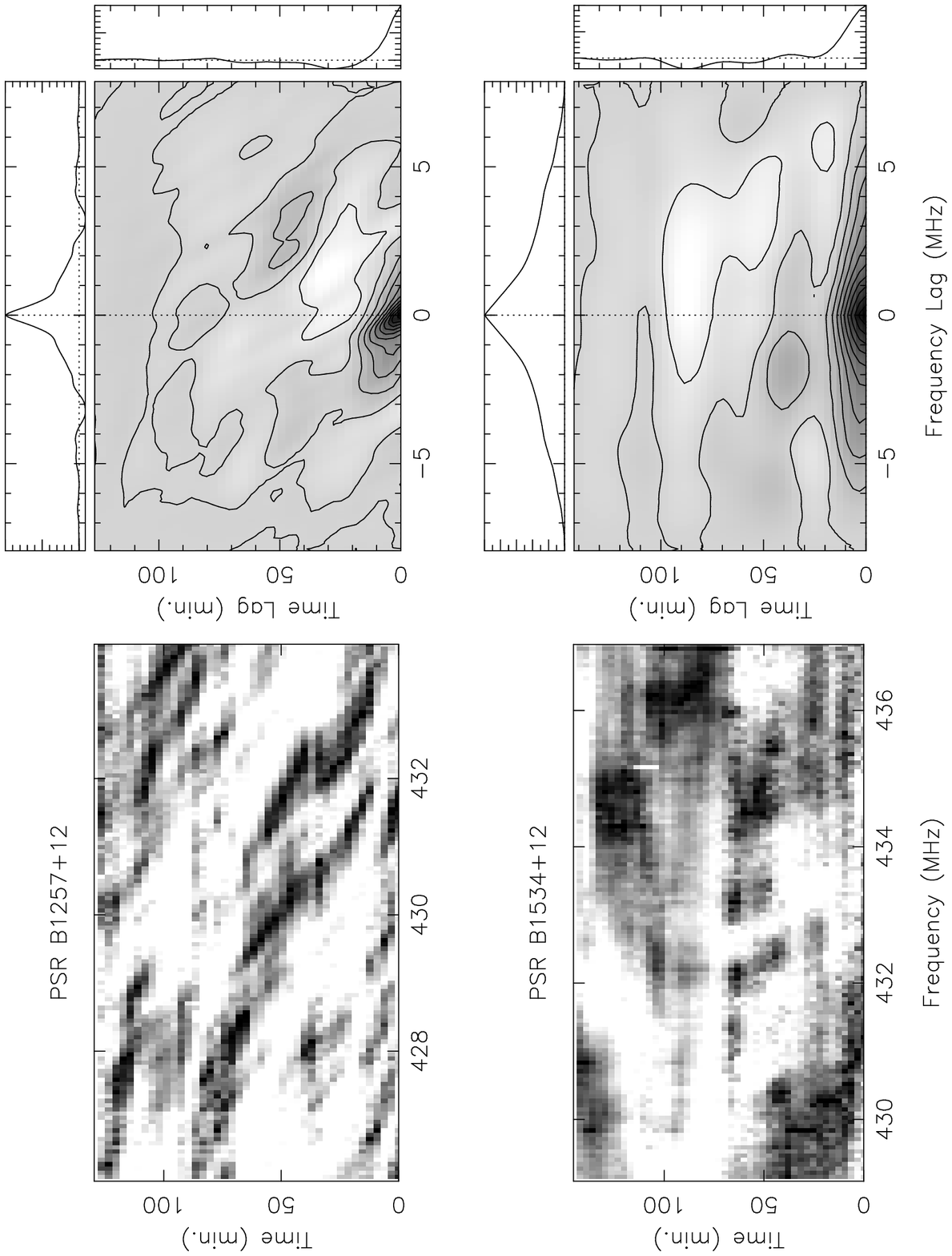]
{{\it (Left)} Dynamic spectra of the interstellar scintillation of PSR B1257+12
and PSR B1534+12 measured with the Arecibo radiotelescope at 430 MHz.
{\it (Right)} The corresponding two-dimensional autocorrelation functions. 
The top and right-hand side panels represent zero lag
cross-sections of these functions in frequency and time, respectively.}

\figcaption[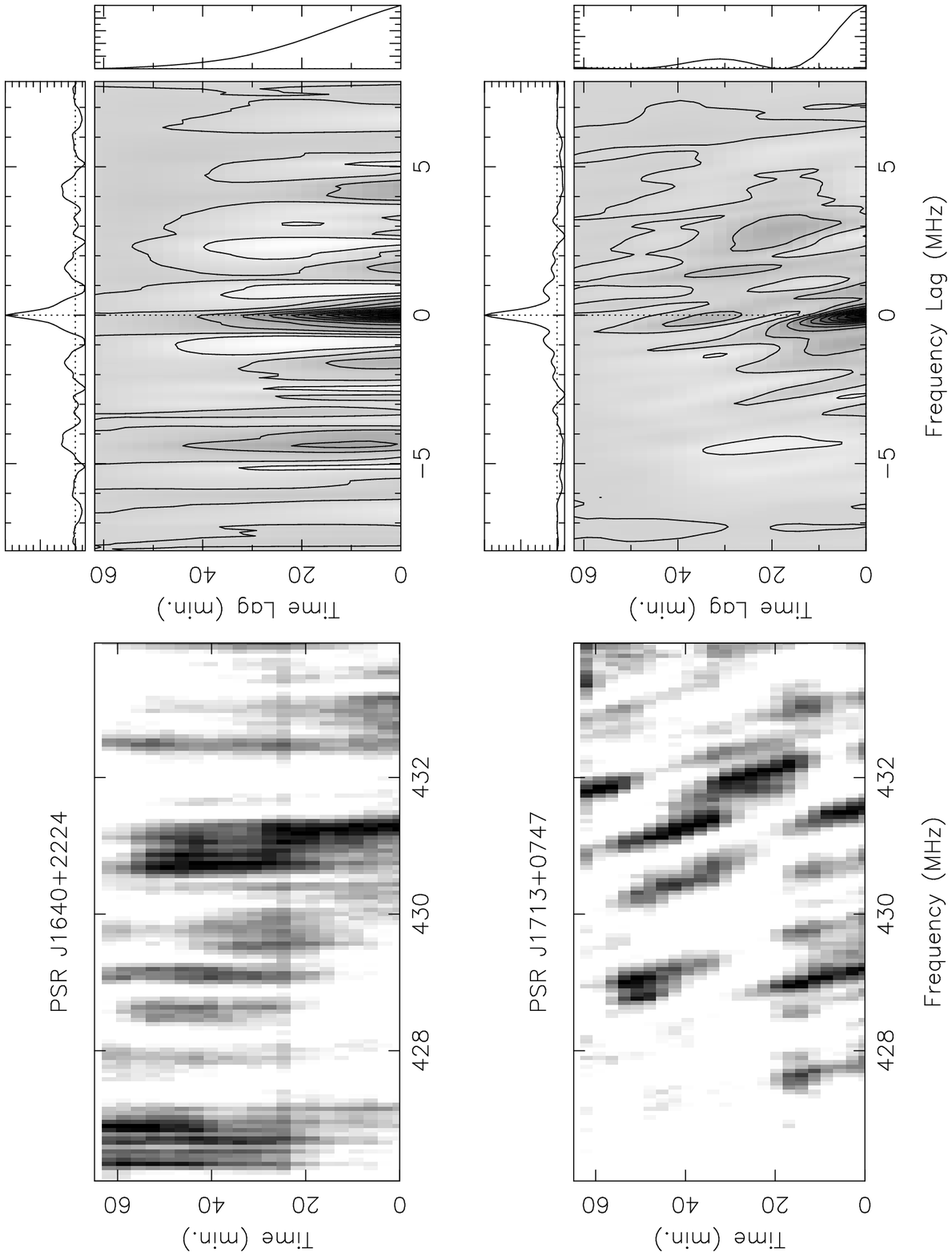]
{{\it (Left)} Dynamic spectra of the interstellar scintillation of 
PSR J1640+2224
and PSR J1713+0747 measured with the Arecibo radiotelescope at 430 MHz.
{\it (Right)} The corresponding two-dimensional autocorrelation functions.
The top and right-hand side panels represent zero lag
cross-sections of these functions in frequency and time, respectively.}

\figcaption[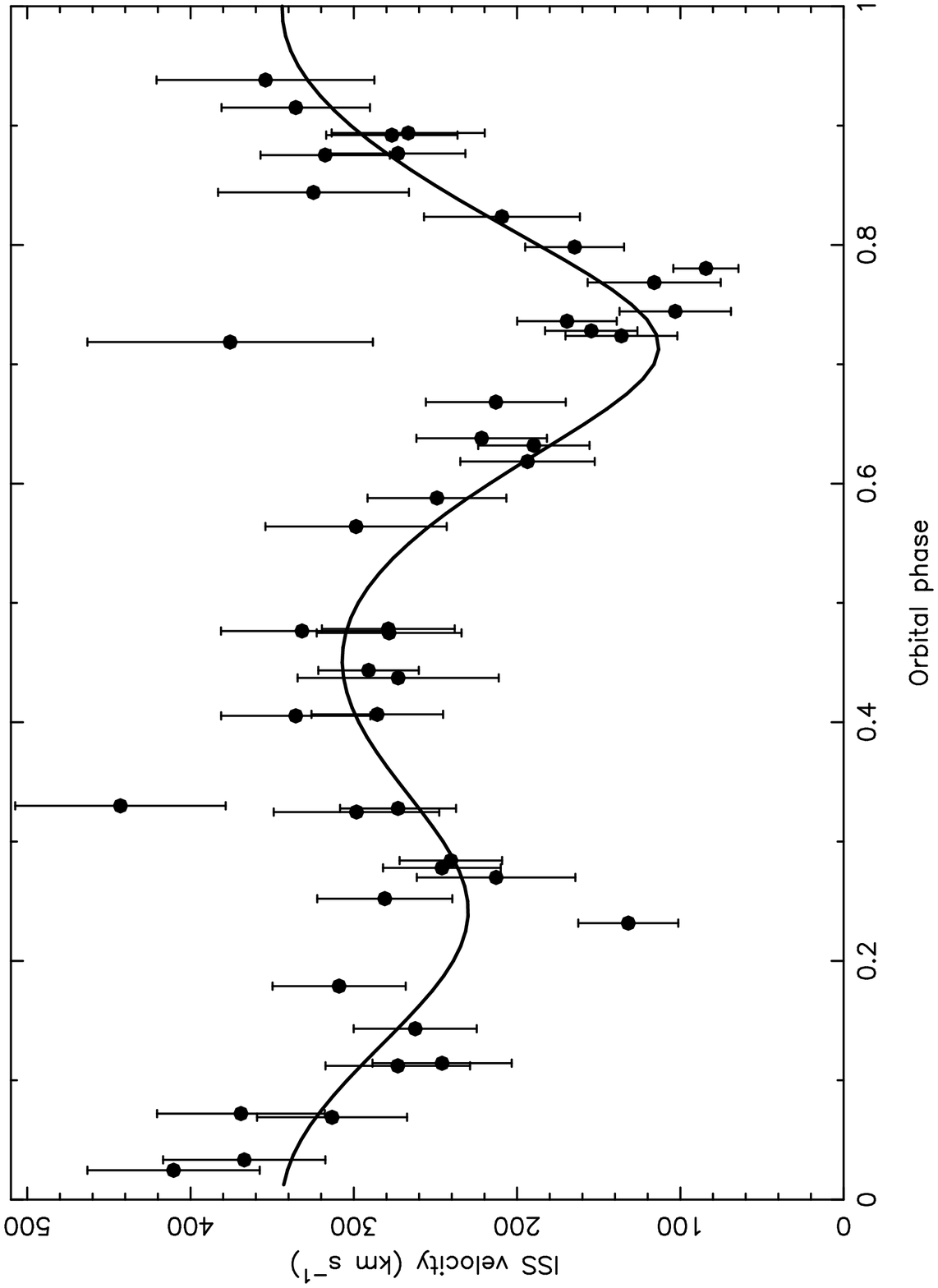]
{The interstellar scintillation velocity of PSR B1534+12 at 430 MHz measured
as a function of orbital phase. The best fit model of this relationship
involving two free parameters, the longitude of the ascending node of the orbit 
$\Omega$, and the ratio of distances of the scattering screen to the observer
and the pulsar $x$, is shown as the solid line.}

\clearpage

%
% TABLES
%

%
% TABLE 1
%

\begin{deluxetable}{lccccc}
\tabletypesize{\small}
\tablecaption{Scintillation parameters and velocities of the pulsars
B1257+12, B1534+12, J1640+2224, and 1713+0747.}

\tablehead{
\colhead{ }  & \colhead{d} & \colhead{$\Delta \nu_d$} & \colhead{$\Delta t_d$} & \colhead{Observed $V_{ISS}$} & \colhead{Predicted $V_{ISS}$} \\ 
\colhead{PSR} & \colhead{(kpc)} & \colhead{(MHz)} & \colhead{(min)} & \colhead{(km s$^{-1}$)} & \colhead{(km s$^{-1}$)} }

\startdata
B1257+12 & 0.6 & $1.1 \pm 0.8$ & $6 \pm 2$  & $197 \pm 57$ & $278 \pm 16$ \\
B1534+12 & 1.1  & $1.1 \pm 0.4$ & $11 \pm 3$ & $192 \pm 63$ & $177\pm11$  \\
J1640+2224 & 1.2 & $0.2 \pm 0.1$ & $20 \pm 4$ & $38 \pm 8$ & $75 \pm 14$ \\
J1713+0747 & 1.1 & $0.6 \pm 0.2$ & $14 \pm 5$ & $82 \pm 16$ & $52 \pm 2$ \\  
\enddata

\end{deluxetable}

\clearpage

%
% FIGURES
%

%
% FIGURE 1
%

\begin{figure}
\figurenum{1}
\epsscale{0.8}
\plotone{f1.ps}
\caption{}
\end{figure}

%
% FIGURE 2
%

\begin{figure}
\figurenum{2}
\epsscale{0.8}
\plotone{f2.ps}
\caption{}
\end{figure}

%
% FIGURE 3
%

\begin{figure}
\figurenum{3}
\epsscale{0.75}
\plotone{f3.ps}
\caption{}
\end{figure}
\end{document}